\documentstyle[12pt]{article}

\newcommand{\beq}{\begin{equation}}
\newcommand{\eeq}{\end{equation}}
\newcommand{\bea}{\begin{eqnarray}}
\newcommand{\eea}{\end{eqnarray}}
\newcommand{\bi}{\bibitem}
\newcommand{\p}{\partial}

\renewcommand{\theequation}{\arabic{section}.\arabic{equation}}
\renewcommand{\thesection}{\arabic{section}.}

\begin{document}

\begin{flushright}
		  gr-qc/9212007
\end{flushright}
\vspace{0.5cm}

\begin{center}
              {\LARGE Weak-Field Gravity  of \\
	      Revolving Circular Cosmic  Strings}
\end{center}

\vskip 1 truecm

\begin{center}{\Large
                     ${}^{\ast}$Des J. Mc Manus}
\end{center}

\begin{center}{\sl
                 School of Theoretical Physics,
                        Dublin Institute for Advanced Studies,\\
                        10 Burlington Road, Dublin 4, Ireland.
              }
\end{center}

 \vskip 5 truemm

{\Large \centerline{${}^{\dag}$Michel A. Vandyck}}

\begin{center}{\sl
                  Physics Department, University College
                        Cork, Ireland, \\ and \\
           \setcounter{equation}{0} 
       Physics Department, Cork Regional Technical College, \\
		  Bishopstown, Co. Cork, Ireland.
               }
\end{center}

\vspace{.5cm}

%

\begin{center}{\large {\bf Abstract}} \end{center}

\noindent
   A weak-field solution of Einstein's equations is
   constructed. It is generated by
   a circular cosmic string revolving in its plane
   about the centre of the circle. (The revolution is
   introduced to prevent the string from collapsing.)
   This solution exhibits a conical singularity, and the corresponding
   deficit angle is the same as for a straight string of the same
   linear energy density, irrespective of the angular velocity
   of the string.

 \noindent PACS numbers:  04.20-q, 98.90+s

\vspace{1cm}

\begin{center}{\it
		  To appear in Physical Review D}
\end{center}
\pagebreak

                          \section{Introduction}                              %

One of  the most notable features of a straight cosmic string
\cite{1} is the presence in spacetime of an angular
deficit, the magnitude of which is related to the linear energy
density $\mu$ of the string by $\delta\psi = 8\pi G \mu$.
Indeed, the deficit-angle structure of spacetime is central to
many proposals for the possible observation of the gravitational effects of
cosmic strings \cite{2}.

 The deficit-angle model is generally accepted as a good approximation
for describing the exterior gravitational field of cosmic strings.
Frolov, Israel, and Unruh (FIU) used precisely this approximation when they
considered
a thin circular string at a moment of time symmetry \cite{FIU}.

Recently, we investigated further \cite{HMV}
the problem of the deficit angle  produced by  circular strings.
We constructed a weak-field stationary
solution of Einstein's equations generated by a thin circular string
and established that external radial stresses had to be introduced to support
the ring
against collapse (thus allowing a stationary solution to exist).
The form of the radial
stresses was completely determined
by stress-energy conservation. The main result of our study
was that a circular string produces conical
singularities with  the same angular deficit  as a straight string of
identical linear energy density, fully supporting FIU's assumption.
Thus, we demonstrated, in the weak-field limit,
the validity of the FIU hypothesis directly from the field equations.
Furthermore, the external radial stresses
were seen not to contribute to this angular deficit.

In the present work, we ask the question whether it is possible to
extend the previous results to a  self-supporting circular string,  as
opposed to an  externally supported circular string: can  the stabilising role
previously played by  external radial stresses be played by an
internal mechanism?   Centrifugal force is the simplest candidate; it should be
possible
to prevent gravitational collapse by spinning the ring at an appropriate
angular velocity.

 We construct a weak-field solution of
Einstein's equations corresponding to an infinitely thin  circular string
revolving at a given  angular velocity $\omega$. We begin by examining the most
general scenario where the string is partially supported by
centrifugal force and
partially by external radial stresses.
The angular velocity is chosen arbitrarily and
the radial stresses
are then determined by stress-energy conservation.
The angular deficit produced by this solution   is found to be  equal
to the deficit produced by a straight string (of the same linear
energy density), irrespective
of the value of the angular velocity $\omega$ (within the
limitation of the weak-field approximation).
Later, we  also calculate and discuss the critical angular
velocity $\omega_{\rm crit}$ at which the ring is totally
self-supporting, namely the particular velocity at which no radial
stresses are necessary to support the ring. The latter is then
supported entirely by centrifugal force.

 In this paper, we use units in which $\hbar = c = 1$, take the metric to have
signature
$(-, +,+,+)$ and adopt the geometrical conventions of Synge \cite {SYNGE}.

%
                       \section{The Stress-Energy Tensor}
%

The stress-energy tensor ${T^\mu}_\nu$ generating
the gravitational field of a  revolving string
partially supported against collapse by external radial
stresses contains three contributions: ${}^{ M}{T^\mu}_\nu$,
the contribution from the circular motion of the ring
(excluding radial stresses required to maintain circular
motion),
${}^{ A}{T^\mu}_{\nu}$, the contribution from the azimuthal
flux through the string (which corresponds to
the ${T^z}_z$ stresses for Vilenkin's straight string),
 and ${}^{ E}{T^\mu}_\nu$, the
contribution from the external radial stresses.
Given that the diameter of the core of a string arising from a
spontaneously broken
gauge theory \cite{PRESKILL} is microscopically small, it is well
justified to make
the approximation that the stress-energy tensors
${}^M{T^\mu}_\nu$ and ${}^A{T^\mu}_\nu$
of the circular string be confined to the infinitely thin ring
$r = a$, $z = 0$, where $a$ denotes the radius of the ring.
On the other hand, the radial stresses ${}^E{T^\mu}_\nu$, being
external, are not confined to the core
of the string. (See  \cite{HMV} for a more detailed discussion.)

For (pressure-free) dust of  rest-energy density $\rho_0$, moving with
four-velocity $u^\mu$,
the stress-energy tensor is
${}^MT^\mu{}_\nu \,=\, \rho_0 \, u^\mu \, u_\nu$, where $u^\mu \, u_\mu = -1$.
In cylindrical coordinates $x^\mu\,\equiv\,(t, \phi, r, z)$, the four-velocity
$u^\mu$ of an infinitely thin ring of radius $a$, centered at
the origin, lying in the
$x$-$y$ plane, and revolving with angular velocity $\omega$ is
readily found as
\bea
    u^\mu \; &=& \; \Gamma \,(\delta^\mu_t \,+\, \omega\,\delta^\mu_\phi)
                                             \label{veloc} \\
    -\Gamma^{-2} \;&\equiv&\; g_{tt}(a,0)\,+\,2\omega \, g_{t\phi}(a,0) \,+\,
        \omega^2\, g_{\phi\phi}(a,0)\;\;\;, \label{gam}
\eea
where $g_{\mu\nu}(a,0)$ denotes the metric components evaluated at
$r=a$, $z=0$.  (The metric enters this equation because of the normalisation
condition $u^\nu \, u_\nu = -1$.)

Furthermore, the ring
has its rest-energy density $\rho_0$ purely localised:
\beq
    \rho_0 \;=\; \mu_0 \; \delta(r-a) \; \delta(z)\;\;\;, \label{rho}
\eeq
where  $\mu_0$ is the linear rest-energy density.
Consequently, the stress-energy tensor ${}^M{T^\mu}_\nu$ produced by the
circular motion has the expression
\beq
    {}^M{T^\mu}_{\nu}\;=\;
    \mu_0\, \Gamma^2 \;
          (\delta^{\mu}_t \,+\,
          \omega\, \delta^{\mu}_\phi)\;
     [g_{\nu t}(a,0) \,+\,  \omega\,
      g_{\nu \phi}(a,0)]\;\delta(r-a)\;\delta(z)\;\;\;,
    \label{CT}
\eeq
where $g$ is the metric and no summation is implied over repeated indices
$t$ and $\phi$. (The summation convention will be suspended all
throughout when $t$, $\phi$ or $r$ appear as subscripts or superscripts.)

The stress-energy tensor giving the contribution from the
azimuthal flux through the string is elementary in the
case of a non-revolving ring \cite{HMV}, the
simple argument being that ${}^A{T^\phi}_\phi$ plays
the role that ${T^z}_z$ plays for a straight string.
However, a more detailed reasoning
is necessary when the string is revolving since a revolving frame is not
inertial. The appropriate stress-energy tensor ${}^A{T^\mu}_\nu$
can be established by employing  the general method
applicable to anisotropic fluids \cite{SYNGE}. (The \lq\lq
fluid\rq\rq\  must be anisotropic since for
physical reasons we expect a pressure along the $\phi$ axis
but none along the $r$ and $z$ axes.)

The stress-energy tensor  \cite{SYNGE} of a fluid
of rest-energy density $\rho_0$,
moving with a four-velocity $u^\mu$, is
\beq
    T^\mu{}_\nu \;=\; \rho_0 \; u^\mu \, u_\nu \;-\; S^\mu{}_\nu \label{an0}
    \;\;\;,
\eeq
in which the tensor $S^\mu{}_\nu$, the stress tensor,  satisfies
$S^\mu{}_\nu \, u^\nu = 0$. The  eigenvectors
$\lambda^\mu_{(i)}$, assumed space-like, and the corresponding
eigenvalues $\, -P_{(i)}, \,1 \leq i \leq 3$ are called
respectively the ``principal axes'' and the ``principal stresses''
of $S^\mu{}_\nu$.
In terms of these quantities, the stress tensor can be expressed as
\beq
    S^\mu{}_\nu \;=\;
         -\sum_{i=1}^3 P_{(i)}\;\lambda^\mu_{(i)}\;
            \lambda_{(i)\nu} \label{an1} \;\;\;,
\eeq
where $\lambda^\mu_{(i)}$  satisfies
\bea
    u_\mu \; \lambda^\mu_{(i)} \;&=&\; 0 \label{an3} \\
    \lambda^\mu_{(i)} \; \lambda_{(j)\mu} \;&=&\;
    \delta_{ij}\;\;\;. \label{an4}
\eea

The stress-energy tensor ${}^MT^\mu{}_\nu$ of (\ref{CT}) already
takes into account the motion of the ring; therefore, we only
consider the stress part $S^\mu{}_\nu$ of (\ref{an0}).
The physical requirement that there be  no
pressure  along the $r$ and $z$ axes implies that the principal pressures
$P_{(2)}$ and $P_{(3)}$ vanish.
Thus, the only non-zero principal pressure is the azimuthal
pressure $P_{(1)} \equiv P_{(\phi)}$. For an infinitely thin ring, the
azimuthal pressure  is confined to the core of the
string, which implies
\beq
    P_{(1)} \;=\; k_0 \;\delta(r-a) \;\delta(z)\;\;\;,
\eeq
where $k_0$ is a constant with dimensions of force. (We interpret
$-k_0$ as the tension in the ring.)

Moreover, the only relevant principal axis, $\lambda_{(1)}$, must
point spatially along the $\phi$ direction.  The constraint (\ref{an4}),
together with (\ref{an3}) applied to the velocity (\ref{veloc}),
determines $\lambda_{(1)}$  as
\beq
    \lambda^\mu_{(1)} \;=\; K \,\delta^\mu_t \,+\, L \,\delta^\mu_\phi
                    \;\;\; ,  \label{lam1}
\eeq
where $K$ and $L$ are solutions of the system
\bea
    K\, g_{tt}(a,0) \,+\, (L+\omega K)g_{t\phi}(a,0)
        \,+\,   \omega L\, g_{\phi\phi}(a,0)\;&=&\;0 \label{lam2} \\
    K^2\, g_{tt}(a,0) \,+\, 2 K L\,  g_{t\phi}(a,0) \,+\,
        L^2\, g_{\phi\phi}(a,0) \;&=&\; 1 \;\;\;. \label{lam3}
\eea
Thus, we reach the conclusion that the stress-energy tensor
${}^A{T^\mu}_{\nu}$ generated by the azimuthal flux along the
string reads
\beq
    {}^A{T^\mu}_\nu \;=\; k_0 \,(K \, \delta^\mu_t \,+\,
    L \, \delta^\mu_\phi) \; [K\, g_{\nu t}(a,0)
    \,+\, L \,g_{\nu \phi}(a,0)] \;\delta(r-a) \;
    \delta(z) \label{flux}
\eeq
(with no summation on $t$ or $\phi$).
This tensor depends on two scalar parameters: $\omega$, the angular velocity
of the ring, and  $-k_0$, the tension in the ring. When $\omega$ vanishes,
(\ref{flux}) becomes identical to the expression obtained in the
non-revolving case \cite{HMV}, so that it is consistent to identify
the scalar $k_0$ with the parameter $k$ in \cite{HMV}. Consequently,
for string matter, it is reasonable to generalise the equation of state
$k = - \mu$ of the non-revolving case  as $k_0 = - \mu_0$
in the revolving case.

In order to support the ring partially by external
radial stresses, we assume the existence of  a radial component of
${}^E{T^\mu}_\nu$  confined to the
$x$-$y$ plane (as in the case \cite{HMV} of a non-revolving ring):
\beq
    {}^E{T^\mu}_\nu \;\equiv\; \Delta(r,z)\;
    \delta^\mu_r \; \delta_\nu^r \;\equiv\; f(r) \; \delta(z)\;
    \delta^\mu_r \; \delta_\nu^r \;\;\;, \label{RT}
\eeq
where $f$ is a function to be determined later
by stress-energy conservation, and there is no summation over $r$.

The complete stress-energy tensor ${T^\mu}_\nu$ generating the
gravitational field of the revolving string
is the sum of all the above contributions (\ref{CT}), (\ref{flux}),
and (\ref{RT}):
\beq
    {T^\mu}_{\nu} \;=\; {}^M{T^\mu}_\nu \,+\, {}^A{T^\mu}_\nu \,+\,
    {}^E{T^\mu}_\nu \;\;\;.\label{TTOTAL}
\eeq
We now turn to the form of the spacetime metric $g$.

\setcounter{equation}{0} 

%
                         \section{The Metric and Field Equations}
%

The most general stationary metric
produced by an axially-symmetric revolving
source
may be written \cite{C&F}
\beq
    ds^2 \;=\; -e^{2\nu}\,dt^2 \;+\;  e^{2\zeta - 2\nu}\, r^2\,
    (d\phi \,-\,Adt)^2\;+\;
    e^{2\eta - 2\nu} \,(dr^2\,+\, dz^2)\;\;\;,
    \label{2_1}
\eeq
where $\nu, \zeta$, $\eta$, and $A$
are functions of $r$ and $z$ only.

We proceed as in the
non-revolving case \cite{HMV} and make the weak-field
approximation. Thus, we calculate the Einstein tensor $G^\mu{}_\nu$
\cite{C&F} for the
metric (\ref{2_1})
and retain only
first-order terms in $\nu$,
$\eta$, $\zeta$,
and  $A$.

Given that $A$ is not dimensionless, it is not immediately apparent
that terms of order $A^2$ may be neglected. However, for the stress-energy
tensor
(\ref{CT}), (\ref{flux})--(\ref{TTOTAL}), the linearised Einstein
field equations
show that this is indeed the case  at first order in the dimensionless
quantities $G \mu_0$ and $G k_0$. (A complete discussion is presented in
Appendix A.) Therefore, in all our future considerations, the expression
``weak-field approximation'' will refer to the expansion of the
field equations at first order in $\nu,\, \eta,\, \zeta,\, A, \,G \mu_0,$ and
$G k_0$.

With this approximation, the equations (\ref{gam}),
(\ref{CT}), and (\ref{lam2})--(\ref{flux}) simplify
greatly, and after some
manipulations
the field equations (see appendix A)
reduce to
\bea
    \nabla^2 \nu &=& 4\pi G \,\{(C_1 \,+\,C_2)
       \, \delta(r-a)\;\delta(z)
        \,+\, \Delta(r,z)\}\label{3_1} \\
    {\widetilde{\nabla}}^2\, \eta &=& 8\pi G \, C_2
               \, \delta(r-a)\;\delta(z)  \label{3_2} \\
    \nabla^2\,\zeta \,+\, \frac{1}{r}\, \zeta_r
               \;&=&\;
              8\pi G\,\Delta(r, z)
                   \label{3_3}\\
    \frac{\p}{\p r} (r \zeta) \;&=&\; \eta - \eta_0 \label{3_4} \\
    \nabla^2 A \,+\, \frac{2}{r}\, A_r \;&=&\;  -16 \pi G \, C_3 \,
           a^{-1}\, \delta(r-a)\;\delta(z) \;\;\;,
           \label{3_5}
\eea
in which
$\Delta$ is the radial-stress function
defined in (\ref{RT}),
$\widetilde{\nabla} \equiv  \p^2_r +
\p^2_z,\; \nabla \equiv
\widetilde{\nabla} + (1/r)\,\p_r$ is the
Laplacian,
$\eta_0$ is an arbitrary constant of integration,
and the constants $C_i$,
$1 \leq i\leq 3$ are related to the parameters
$\mu_0$, $k_0$, $\omega$, and $a$ of the problem by
\bea
    C_1 \;&\equiv&\;(\mu_0\, +\,
      \omega^2 a^2\;k_0)\Gamma^2_0   \label{k1} \\
    C_2 \;&\equiv&\;  (\omega^2 a^2 \; \mu_0 \,+\, k_0)
           \Gamma^2_0    \label{k2} \\
    C_3 \;&\equiv&\; (\mu_0\,+\, k_0) \Gamma^2_0 \; \omega a
                  \label{k3} \\
    \Gamma^{-2}_0 \;&\equiv&\;  1\, - \, \omega^2 \, a^2
      \;\;\;\;.\label{k4}
\eea
(The quantities $G C_i, \; 1 \leq i \leq 3$, are dimensionless.)

The compatibility condition for
 (\ref{3_2})--(\ref{3_4}), or
equivalently the stress-energy conservation law,  determines
the radial-stress function $\Delta$ of (\ref{RT}) as
\beq
    \Delta(r, z) \;=\; f(r)\;\delta(z) \;=\; (C_2/r)
                 \; \Theta(r-a) \;\delta(z)
     \;\;\;,\label{3_9}
\eeq
in which $\Theta$  denotes the Heaviside
step function.  This expression, which is similar to the one obtained
in the non-revolving case \cite{HMV}, may now
be substituted into (\ref{3_1}), so that we have
a complete set of equations of which the metric
functions $\nu$, $\zeta$, $\eta$, and $A$ are
solutions.

\setcounter{equation}{0} 

%
		\section{Angular Deficit}
%

Our main purpose in solving  the field equations
(\ref{3_1})--(\ref{3_5}) is to
investigate the metric for conical singularities and to
calculate the corresponding angular deficit.
Because of the fact that conical singularities involve only
the metric (\ref{2_1})
at constant time $t$ and constant azimuth
$\phi$, it is sufficient to restrict attention to
obtaining explicitly the
functions $\nu$ and $\eta$ given by (\ref{3_1}) and
(\ref{3_2}). We solved these equations, for different
values of the constants $C_i$, in our
previous work on the non-revolving ring \cite{HMV}, and it
is therefore not necessary to repeat the analysis here.  We
only recall that, in toroidal \cite{BATEMAN} coordinates
$(t, \phi, \sigma, \psi)$, which are related to the
cylindrical coordinates $(t, \phi, r, z)$ by
\beq
    z/a\;\equiv\; N^{-2} \sin\psi \qquad
    r/a\;\equiv\; N^{-2} {\rm sh}\,\sigma \qquad
    N^2 \;\equiv\;N^2(\sigma,\psi) \;\equiv\;
      {\rm ch}\,\sigma \,-\, \cos\psi\;\;\;,  \label{toroidals}
\eeq
the solutions near the string (namely for $\sigma \rightarrow
\infty$) read:
\bea
    \nu(\sigma,\psi) \;&\rightarrow&\;  -2 G \, (C_1 \,+\, C_2)
     \,\sigma \label{nunear} \\
     \eta(\sigma,\psi) \;&\rightarrow&\; -4 G \, C_2
      \,\sigma \;\;\;.\label{etanear}
\eea
(As in the non-revolving case \cite{HMV}, the radial
stresses (\ref{RT}) and (\ref{3_9}) do not contribute to
these asymptotic forms for $\nu$ and $\eta$, and thus have
no influence on the conical singularities.)

It follows from (\ref{nunear}) and (\ref{etanear})
that the combination $\eta - \nu$, which determines
the metric (\ref{2_1})  at constant $t$ and $\phi$, becomes,
after substituting the definitions (\ref{k1}), (\ref{k2}),
(\ref{k4}) and noting the  non-trivial cancellation of the
$\omega$-dependent terms:
\beq
    \eta - \nu \;\rightarrow\; 2G(\mu_0 \,-\, k_0)\sigma   \;\;\;.
     \label{muk}
\eeq
The fact that $\eta - \nu$ is proportional to the toroidal
coordinate $\sigma$ indicates the presence of a conical
singularity \cite{FIU,HMV}. The corresponding
angular deficit $\delta\psi$, which is related to the  ratio
of the perimeter of a circle centered at the core of the string
to the radius of this circle  \cite{FIU,HMV}, is given by
\beq
    \delta\psi \;=\; 2\pi \;-\;
    \lim_{\sigma_0\rightarrow\infty}
    \left[\int_{-\pi}^{\pi} d\psi \, (N^{-2} \,
    {\rm e}^{\eta-\nu})
    \mid_{\sigma = \sigma_0} \left/
    \int_{\sigma_0}^\infty d\sigma\, N^{-2} \, {\rm e}^{\eta-\nu}
    \right. \right] \;\;\;,\label{integrall}
\eeq
and is easily calculated as
\beq
    \delta \psi \;=\; 4\pi G (\mu_0 \,-\, k_0)\;\;\;.\label{fin}
\eeq

As announced earlier, this expression  is independent of the angular
velocity $\omega$ at which the ring revolves. Moreover, the angular
deficit is also identical with Vilenkin's result for
a straight string.
(For string matter, $k_0 = -\mu_0$, as explained in \cite{HMV}.)
We have thus  demonstrated, in the weak-field approximation, that
a revolving circular string produces the same angular deficit
 as a straight
string of the same linear energy density.

\setcounter{equation}{0} 

%
	     \section{Self-supporting Ring}
%

Finally, we address  the problem  of whether a revolving string
can be totally self-supporting. Up to this point, we considered a ring
partially supported by the Centrifugal Force produced by
the revolution and partially supported by the external radial
stresses $\Delta$ of (\ref{RT}) and (\ref{3_9}).

For the discussion that follows, the equation of state relating
$k_0$ and $\mu_0$ will conveniently be written as
\beq
    k_0 \,=\, (\alpha - 1)\, \mu_0 \;\;\;, \label{5_1}
\eeq
where $\alpha$ is a parameter. ( This particular parametrisation
excludes the possibility of the physically uninteresting case
$\mu_0 = 0, k_0 \neq 0$.) String matter is characterised by $\alpha = 0$,
whereas non-string matter has negative pressure for $0 < \alpha < 1$ and
positive pressure if $\alpha > 1$.

By definition the ring is self-supporting when no external radial stresses
are necessary to prevent collapse.
This happens, by virtue of (\ref{3_9}), if and only if
$\omega$ takes the critical value $\omega_{\rm crit}$
that forces $C_2$ to vanish. Equations (\ref{k2}) and (\ref{k4}), upon
inserting (\ref{5_1}), imply that $\omega_{\rm crit}$ is the solution
of
\beq
    0 \,=\, \mu_0 \, \left[ -1 \,+\,
    \frac{\alpha}{1 - (\omega_{\rm crit} a)^2} \right] \;\;\;. \label{5_2}
\eeq
The above constraint always has the trivial solution $\mu_0 = 0$, which in
turn implies $k_0 = 0$, so that spacetime is flat everywhere. For string
matter $\alpha = 0$, and  this
trivial solution is the only solution that (\ref{5_2}) admits.
Consequently, we have established that a ring of string matter cannot
be made self-supporting exclusively by centrifugal force but that a
certain amount of external stress is necessary to prevent collapse.
In other words, trying to support a ring of string matter purely by
inducing a revolution requires the string to be massless, which is
non physical.

The solution of (\ref{5_2}) in the physically interesting case,
$\mu_0 \neq 0$, is
\beq
    (\omega_{\rm crit} a)^2 \,=\, 1 \,-\, \alpha \,\,\,. \label{5_3}
\eeq
 We observe that $\omega_{\rm crit}$ is independent of $G$, and therefore
of the gravitational field. This is a simple manifestation of the
well-known \cite{OHAN}
\lq\lq Motion of the Source\rq\rq\ problem in the linearised
Einstein equations:
the linear approximation is sufficient to calculate the first
metric correction produced by the source but neglects the
back-reaction of gravity onto the source, so that the source
moves as is gravity were absent. Taking this back-reaction into
account requires using at least second-order terms as done, for
instance, in the Einstein-Infeld-Hoffman procedure \cite{EIH} for
the motion of point-masses.
 It is important,
however, to insist on the fact that, although the first-order
framework is not appropriate to study gravitational influences
on the {\it motion of the source}, it is perfectly valid to study
gravitational corrections to the {\it metric}, and thus our result
on the angular deficit holds.

Further insight on the physical significance of (\ref{5_3}) may be
gained by studying, in Classical Mechanics,  the equilibrium condition
for a revolving ring of radius $a$, linear mass density $\mu$  and
tension $T$. Consider a small arc of angular width
$\theta$ along the circle. The two extremities of this arc are subjected
to a tension $T$ which is tangential, but the resultant force $T_R$ at the
mid-point along the arc is purely radial and is given by
$T_R \,=\, 2 T \sin{(\theta/2)}$. On the other hand, the centrifugal
force $F_C$ acting on the arc reads $F_C \,=\, \mu \theta \omega^2\,
a^2$, and consequently, equilibrium is attained   when $\omega$
reaches the critical value $\omega_{\rm crit}$ satisfying
\beq
    (\omega_{\rm crit}\,a)^2 \;=\;
    \frac{T}{\mu} \, \lim_{\theta \rightarrow 0} [(2/\theta)
     \sin{(\theta/2)}] \;=\; \frac{T}{\mu} \;\;\;.
\eeq
This result is
identical with (\ref{5_3}) for $T = -k_0$ and $\mu = \mu_0$ since
$\alpha = 1 + k_0/\mu_0$ by (\ref{5_1}).

It follows from  (\ref{5_3})
that matter must have a negative pressure to lead
to a self-supporting ring [since $\alpha$ must be less than one for
(\ref{5_3}) to admit a solution for $\omega_{\rm crit}$].
This is physically reasonable
since a positive pressure would tend to create an expansion of the
ring, whereas a negative pressure would tend to create a collapse.
Only a collapse, and therefore
a negative pressure,  could be counteracted by Centrifugal Force.
(The  gravitational force does not contribute to the
collapse at first order, as just mentioned.)

Moreover, as the pressure becomes more negative and approaches
$-\mu_0$,
the critical
velocity increases. The extreme case of
self-supporting string matter, namely  $\alpha = 0$,   formally
implies that $\omega_{\rm crit} a = \pm 1$, which means that the
string
revolves at the speed of light.
This is impossible for  a massive body.
Thus, we reach the same conclusion as before that string matter cannot
be made self-supporting by revolution only.
(This may no longer hold if gravity is explicitly
taken into account in the motion of the source.)
The present argument, albeit physically enlightening, is only
formal since the weak-field approximation breaks down for a
massive body if $\omega a = \pm 1$.

We emphasise that the stress-energy tensor used in our treatment
contains terms proportional to $\mu_0\, (1-a^2\omega^2)^{-1}$ and
$k_0 \, (1-a^2\omega^2)^{-1}$. Therefore, our results are valid
when the dimensionless quantities $G \mu_0\, (1 - a^2\omega^2)^{-1}$
and $G k_0 \, (1 - a^2 \omega^2)^{-1}$ are small. In particular, the
value of $\omega a$, for a ring partially supported by revolution and
partially by external stresses, is not determined by any equation [in
contrast with the totally self-supporting case (\ref{5_2}), (\ref{5_3})],
but is a free parameter.
Therefore, our main result, namely that the angular deficit produced
by a revolving
ring is the same as for a straight string of the same linear density,
is valid, within the confines of the weak-field limit, for a large
range of the values of the parameter $a\omega$.

\setcounter{equation}{0} 

%
		   \section{Conclusion}
%

In this work, we extended our previous results \cite{HMV} on the deficit-angle
structure of spacetime of a non-revolving circular string to the case of
a revolving circular string. The circular string was prevented from collapse
partially by the centrifugal force produced by the revolution and
partially by external radial stresses that were determined by
stress-energy conservation.
We established, in a weak-field treatment,
that a conical singularity exists along the string and
that the magnitude of the corresponding deficit angle is the same as
that produced
by a straight string of the same linear energy density.

We also investigated whether a revolving circular ring could be totally
self-supporting, that is whether there existed a critical angular velocity
$\omega_{\rm crit}$ at which the ring is in equilibrium without the presence
of external radial stresses.
We took string matter to have the equation of state $k_0 = -\mu_0$,
where $-k_0$ and $\mu_0$ denote the tension and the linear rest-energy
of the string, respectively, and found that it was impossible to
have a self-supporting string. However, for non-string matter characterised by
the equation of state $k_0 \neq -\mu_0$, the critical angular velocity was
established to be
\beq
     (\omega_{\rm crit} a)^2 \;=\; -k_0/ \mu_0 \;\;\;,
\eeq
where $a$ is the radius of the ring.
[The weak-field treatment was shown to be valid as long as the
dimensionless quantities $G \mu_0(1 - a^2 \omega^2)^{-1}$ and
$G k_0(1-a^2 \omega^2)^{-1}$ are small compared to unity.]

\section*{Acknowledgements:}

M.V. gratefully acknowledges the Royal Irish Academy for a grant
from the Research Project Development Fund and the Cork Regional
Technical College (in particular Mr. R. Langford and Mr. P. Kelleher)
for Leave of Absence. It is a pleasure to
thank S.J. Hughes and L. O'Raifeartaigh for enlightening
discussions about this problem.

%
%
\setcounter{equation}{0}
\setcounter{section}{0}

\renewcommand{\thesection}{Appendix  A.}
\renewcommand{\theequation}{\Alph{section}.\arabic{equation}}

\section{Frame components of the Field equations}

The most general stationary metric produced by an axially-symmetric
source, namely (\ref{2_1}),
may be written in terms of an orthonormal basis, $\{ \theta^{(\hat{\mu})}\}$
as
\beq
    ds^2 \;=\; \eta_{\mu \nu} \, \theta^{(\hat{\mu})} \, \otimes
    \theta^{(\hat{\nu})} \;\;\;, \label{A1}
\eeq
where $\eta_{\mu \nu} = {\rm diag} (-1, 1, 1, 1)$ and
\bea
    \theta^{(\hat{0})} \;\equiv\; e^\nu \, dt & \;\;\;\;\;\; &
    \theta^{(\hat{1})} \;\equiv\; r e^{\zeta - \nu} \, (d\phi \,-\,  A dt)
    \nonumber \\
    \theta^{(\hat{2})} \;\equiv\; e^{\eta - \nu} \, dr &\;\;\;\;\;\;&
    \theta^{(\hat{3})} \;\equiv\; e^{\eta - \nu} \, dz \;\;\;. \label{A2}
\eea
We define the transformation matrix $e^{\hat{\mu}}{}_\nu$ (and its
inverse $e^\mu{}_{\hat{\nu}}$) by
\beq
    \theta^{(\hat{\mu})} \; \equiv \;  e^{\hat{\mu}}{}_\nu \, dx^\nu \;\;\;.
\eeq
The  components of the
Einstein tensor and the stress-energy tensor in the frame
$\theta^{(\hat{\mu})}$ are then simply
\bea
    G_{\hat{\mu} \hat{\nu}} \;=\; G_{\alpha\beta}\,
    e^{\alpha}{}_{\hat{\mu}} \, e^{\beta}{}_{\hat{\nu}} \;\;\; &;&
    \;\;\;
    T_{\hat{\mu} \hat{\nu}} \;=\; T_{\alpha\beta}\,
    e^{\alpha}{}_{\hat{\mu}} \, e^{\beta}{}_{\hat{\nu}} \;\;\;,
\eea
and Einstein's equations read
$G_{\hat{\mu} \hat{\nu}} = -8\pi G \, T_{\hat{\mu} \hat{\nu}}$.

We now turn to the weak-field approximation.
The metric (\ref{A1}) is flat if $\nu = \eta = \zeta = 0$ and $A$ is
constant. Furthermore, the quantities $\nu, \eta, \zeta$ are dimensionless,
whereas $A$ has dimensions of $1$/length. Therefore, it is not immediately
clear that
a  meaningful approximation to Einstein's equations can be obtained by
neglecting terms of order $A^2$.

To clarify this point, we begin by calculating the Einstein tensor
in the orthonormal frame (\ref{A2}) up to terms of first order in
$\nu, \eta, \zeta$
and all orders in $A$.
The results are
\bea
     G_{\hat{0}\hat{0}} &=&  -2(\nabla^2 \nu) \,+\, (\nabla^2 \,+\,
     \frac{1}{r}\,\p_r) \zeta \,+\, \widetilde{\nabla}^2 \eta \,+\,
     \frac{r^2}{4}  \,(1+2\epsilon)\, (A^2_r \,+\, A^2_z) \label{G00}  \\
     G_{\hat{1}\hat{1}} &=&
     - \widetilde{\nabla}^2 \eta  \,+\,
     \frac{3 r^2}{4}  \,(1+2\epsilon)\,(A^2_r \,+\, A^2_z) \label{G11} \\
     G_{\hat{2}\hat{2}} &=& -\zeta_{zz}  \,-\, \frac{\eta_r}{r} \, +\,
     \frac{r^2}{4}\,(1+2\epsilon)  \,(A^2_z \,-\, A^2_r) \label{G22} \\
     G_{\hat{3}\hat{3}} &=& -\zeta_{rr} \,+\, \frac{1}{r}(\eta_r \,-\, 2
     \zeta_r) \,+\, \frac{r^2}{4}\, (1+2\epsilon)\,
     (A^2_r \,-\, A^2_z) \label{G33} \\
     G_{\hat{2}\hat{3}} &=& \zeta_{rz} \,+\, \frac{1}{r}\, (\zeta_z
     \,-\, \eta_z) \,-\, \frac{r^2}{2} \,(1+2\epsilon)
     \, A_r\, A_z \label{G23} \\
     G_{\hat{0}\hat{1}} &=& -\frac{r}{2}\,{\cal L}A \label{G01} \;\;\;,
\eea
with
\beq
    {\cal L} \;\equiv\; e^{-2(\eta + \zeta - 2\nu)} \left\{ \p_r
    \left[ e^{(3\zeta - 4 \nu)}\, \p_r \right] \,+\,
    \p_z \left[ e^{(3\zeta - 4 \nu)} \, \p_z \right] \,+\,
    \frac{3}{r}\, \left[ e^{(3\zeta - 4\nu)} \, \p_r \right] \right\}
    \label{A11} \;\;\;,
\eeq
where $\epsilon \equiv \zeta - \nu - \eta, \,
\widetilde{\nabla}^2 \equiv \p^2_r + \p^2_z$ and
$\nabla^2 \equiv \widetilde{\nabla}^2 + (1/r)\,\p_r$ is the Laplacian
operator. [In equation (\ref{A11}), the exponentials are only to be taken to
first order in their arguments.]

The stress-energy tensors (\ref{CT}) and (\ref{flux}) are proportional to
$\mu_0$ and $k_0$, respectively, and thus, their contributions to the
Einstein equations are proportional to the dimensionless quantities
$G \mu_0$ and $G k_0$. Therefore, the metric appearing in the expressions
for the stress-energy tensors (\ref{CT}) and (\ref{flux})
may be replaced by the
flat-space metric, namely $\nu = \eta = \zeta = 0, A=$const, if
attention is restricted to a
first-order treatment in $G \mu_0$ and $G k_0$.
Hence, without making any approximation on the order of $A$,
we calculate that the $T_{\hat{0}\hat{1}}$ component of
the total stress-energy tensor (\ref{TTOTAL})  is
\beq
    G\, T_{\hat{0}\hat{1}} \,=\, G (\mu_0 \,+\, k_0) \;
    \frac{a(A - \omega)}{1 - a^2 (A - \omega)^2} \,\delta(r-a)\, \delta(z)
    \;\;\;. \label{T01}
\eeq
Consequently, the Einstein equation
$G_{\hat{0}\hat{1}} = -8\pi G T_{\hat{0}\hat{1}}$ becomes
\beq
    {\cal L} A \;=\; 16 \pi G(\mu_0 \,+\, k_0) \;
    \frac{A - \omega}{1 - a^2 (A - \omega)^2} \,\delta(r-a)\, \delta(z)
    \;\;\;. \label{A_13}
\eeq

We now introduce the dimensionless function $B$ by
\beq
    A \;\equiv\; -16\pi G(\mu_0 \,+\, k_0)\, \omega B \label{AB} \;\;\;.
\eeq
Note that $A$ tends to zero with $\omega$, so that the above parametrisation
guarantees compatibility between the revolving case and the non-revolving
case \cite{HMV},
where $A$ vanishes identically. Thus, the following equation for $B$
is equivalent to (\ref{A_13}):
\beq
    {\cal L} B \;=\; \frac{1\,+\, 16 \pi G(\mu_0\,+\,k_0)\, \omega B}{1
	 \,-\, a^2\omega^2[1\,+\, 16 \pi G (\mu_0\,+\,k_0)B]^2}
	 \; \delta(r-a)\, \delta(z)\;\;\;. \label{A_15}
\eeq

Expanding $B$ in powers of the dimensionless quantity $G(\mu_0 \,+\, k_0)$,
we clearly see from (\ref{A_15}) that the leading term ${}^0\!B$ is of
zeroth order in $G(\mu_0 +k_0)$ and satisfies
\beq
    {\cal L}\, {}^0\!B \;=\; \frac{1}{1-a^2\omega^2}\, \delta(r-a)\, \delta(z)
    \;\;\;.
\eeq
As a result, by virtue of (\ref{AB}),
the leading term in $A$ is proportional to $G(\mu_0 + k_0)$.
Therefore, we may neglect second powers of $A$ and products of $A$ with
$\nu, \eta$, or $\zeta$ in our approximation scheme, since we only
retain first powers of $G \mu_0$ and $G k_0$.

\rule{13.3cm}{0.5pt}

\begin{tabbing}
       ${}^{\ast}$\= \kill
       ${}^{\ast}$Address after September 1, 1992: Dept. of
       Mathematics, Statistics, and Computer \\
        \>Science, Dalhousie University,
       Halifax, Nova Scotia, Canada B3H 4H6. \\
       ${}^{\dag}$Research Associate of the Dublin Institute for
                 Advanced Studies.
\end{tabbing}


\end{document}